\documentclass[12pt]{iopart}
\usepackage{graphicx}

\begin{document}
\jl{2}

\title[Elementary excitation families and their frequency ordering in
       cylindrical BECs]
      {Elementary excitation families and their frequency ordering in 
       cylindrically symmetric Bose--Einstein condensates}
\author{A A Penckwitt and R J Ballagh}

\address{Department of Physics, University of Otago, Dunedin, NZ}

\begin{abstract}
We present a systematic classification of the elementary excitations of
Bose--Einstein condensates in cylindrical traps in terms of their shapes. The
classification generalizes the concept of \emph{families} of excitations first
identified by Hutchinson and Zaremba (1998) \emph{Phys. Rev.} A \textbf{57}
1280 by introducing a second classification number that allows all possible
modes to be assigned to a family. We relate the energy ordering of the modes to
their family classification, and provide a simple model which explains the
relationship.
\end{abstract}

\section{Introduction}

Collective modes provide important signatures of the physics of Bose--Einstein
condensates, and they have been intensively studied over the past few years. A
large number of theoretical calculations, both numerical (e.g.\
\cite{Edwards96}--\cite{Cerboneschi98}) and analytical (e.g.\
\cite{Fetter96}--\cite{Braaten99}), have been made based on the standard
Bogoliubov treatment, and experimental observations \cite{Jin96}--\cite{Fort00}
have confirmed that at very cold temperatures the Bogoliubov--de Gennes (BdG)
equations provide a very accurate description of these modes. This method
amounts essentially to specifying eigenmodes for a perturbative, linearized set
of equations around a potential formed by the trap and the ground state
condensate wave function. In this paper we use the standard BdG equations to
make a systematic study of the shape of elementary modes and the relationship
to their energy ordering. We note that very recent studies have incorporated
finite temperature effects in order to improve accuracy at temperatures closer
to the critical temperature (e.g.\ see \cite{Fedichev98}--\cite{Giorgini00} and
references therein), however, for our purposes, the BdG equations provide an
appropriate and tractable formalism.

For an isotropic trap, the spherical symmetry allows a separation into a radial
wave function and spherical harmonics, thus much of the behaviour can be
predicted from familiar cases in linear quantum mechanics. For axi-symmetric
traps, which is the case for most experimental situations, such a separation is
not possible, and a variety of techniques have been used to obtain the mode
functions. In an elegant paper, Fliesser \textit{et al}  \cite{Fliesser97}
obtained approximate analytic solutions in the hydrodynamic limit by
identifying a non-trivial operator which commutes with both the eigenvalue
operator of the BdG equations, and the angular momentum component along the
symmetry axis. Hutchinson and Zaremba \cite{Hutchinson98} obtained numerical
solutions for the low-lying modes of axially symmetric traps, and studied the
behaviour of these modes as the asymmetry of the trap was changed from prolate
to oblate. The latter authors noticed that the lowest lying modes could be
grouped into four distinct families, based on the behaviour of the mode
frequency as a function of the trap asymmetry. They also showed by example how
some of their modes could be identified with the modes found by Fliesser
\textit{et al}.

In this paper, we extend the results of Hutchinson and Zaremba, and Fliesser 
\textit{et al} by obtaining a classification of the modes in terms of their
geometrical shapes. This enables us to give a generalized definition of the
mode families (which are in principle of unlimited number). Our classification
into families can be expressed in terms of well defined quantum numbers, which
can be directly related to the quantum numbers used by Fliesser \textit{et al}.
In addition, we have examined the energy ordering of the modes for the case of
oblate and prolate traps, and give a simple model that relates the
characteristic quasiparticle shapes to their energy ordering. This also explains
the results of Hutchinson and Zaremba, that all modes in a family have the same
frequency dependence on trap anisotropy.

\section{Formalism}

The ground state of a dilute Bose--Einstein condensate at low temperatures is
well described by the Gross--Pitaevskii equation (GPE) 
\begin{equation}
  \left[\hat{H}_{0}+NU_{0}|\psi (\bi{r})|^{2}\right] \psi (\bi{r})
    = \mu \psi (\bi{r}),  
  \label{statGPE}
\end{equation}
where 
\begin{equation}
  \hat{H}_{0}=-\frac{\hbar ^{2}}{2m}\nabla ^{2}+V_{\mathrm{trap}}(\bi{r})
  \label{singpartHam}
\end{equation}
is the single-particle Hamiltonian of the trap and $N$ the number of atoms. The
effective interaction strength $U_{0}$ is related to the s-wave scattering
length $a$ and the atomic mass $m$ by $U_{0}=4\pi \hbar ^{2}a/m$ (e.g.\ see
\cite{Dalfovo98}). The wave function $\psi(\bi{r})$ is normalized to unity 
according to $\int |\psi (\bi{r})|^{2}\rmd\bi{r}=1$ and the eigenvalue $\mu $ 
is the chemical potential of the condensate. The quasiparticle excitations on 
the ground state $\psi _{\mathrm{g}}(\bi{r})$ have amplitudes $u_{i}$ and 
$v_{i}$ determined by the Bogoliubov--de Gennes (BdG) equations 
\begin{equation}
  \eqalign{ \mathcal{L}u_{i}(\bi{r})+U_{0}\psi_{\mathrm{g}}^{2} (\bi{r})
              v_{i}(\bi{r}) & = \hbar \omega _{i}u_{i}(\bi{r})\\
            \mathcal{L}v_{i}(\bi{r})+U_{0}\psi_{\mathrm{g}}^{*2} (\bi{r})
              u_{i}(\bi{r}) & = -\hbar \omega _{i}v_{i}(\bi{r}),}  
  \label{BdG}
\end{equation}
where $\mathcal{L}=\hat{H}_{0}+2NU_{0}|\psi _{\mathrm{g}}(\bi{r})|^{2}-\mu$
\cite{Edwards96}. We note that the solutions $u_{i}$ and $v_{i}$ of equations
(\ref{BdG}) are not necessarily orthogonal to the ground state, although this
is required in a fully quantum mechanical treatment. However, since the
orthogonal excitations can readily be obtained from the solutions of equation
(\ref{BdG}) by projection \cite{Morgan98} we will present only the direct
solutions of equations (\ref{BdG}) here.

In this paper we will consider the excitations on an anisotropic ground
state of a condensate in a cylindrically symmetric harmonic trap with
trapping potential 
\begin{equation}
  V_{\mathrm{trap}}(\bi{r})=\frac{1}{2}m\omega _\mathrm{r}^{2}
  [x^{2}+y^{2}+(\lambda z)^{2}],
\end{equation}
where $\lambda =\omega _\mathrm{z}/\omega _\mathrm{r}$ is the anisotropy
parameter of the trap. Our numerical solutions for the ground state of the GPE
(\ref{statGPE}) and for the corresponding BdG equations (\ref{BdG}) are
expressed in units of the harmonic oscillator length $ r_0=\sqrt{\hbar
/2m\omega _\mathrm{r}}$ and energy $\hbar \omega _\mathrm{r}$. We will
illustrate our discussions using the case of $\omega _\mathrm{r}=2\pi \cdot
75$~Hz and non-linearity parameter $C=NU_{0}/\hbar \omega _\mathrm{r}r_0^3=
332$ (unless otherwise stated) though we have tested our results for a wide
range of $C$ values of up to 300000. Because the wave function $v_{i}$ is in
general fairly much a mirror image of $u_{i}$, and $|u_{i}|\gg |v_{i}|$ (apart
from the lowest excitations) \cite{You97}, we present only the wave function
$u_{i}$.

\section{Symmetries}

\label{sec:trapsym}

For isotropic harmonic traps the ground state of the GPE has $l=0$, and so the
GPE is completely separable and reduces to a one-dimensional problem in the
radial coordinate $r.$ The BdG equations (for excitations on the ground state)
are therefore also spherically symmetric, and thus the total angular momentum
and the $z$-component of the angular momentum are conserved. This allows a
separation of the angular dependence in the usual fashion by writing the
quasiparticle amplitudes as $u(\bi{r})=u_{\mathrm{r}}(r)Y_{lm}(\vartheta
,\varphi )$ and the equivalent for $v(\bi{r})$, where $Y_{lm}(\vartheta
,\varphi )$ are spherical harmonics. Since the radial equation for the
excitations does not contain any dependence on $m$, the solutions with the same
$l$ and different $m$ are degenerate.

For cylindrically symmetric harmonic traps a complete separation of variables
is not possible. Although the ordinary Schr\"{o}dinger equation is separable in
cartesian or cylindrical coordinates \cite{Eisenhart48}, this separation is not
possible for the GPE due to the non-linear term. Nevertheless, solutions of the
form 
\begin{equation}
  \psi (\bi{r})=\psi(\rho,z)\rme^{\rmi m_\mathrm{c}\varphi }  
  \label{separation}
\end{equation}
can be found, where $\rho $, $\varphi $ and $z$ denote the usual cylindrical
coordinates. Then, neither the trapping potential nor $|\psi (\bi{r})|^{2}$ in
the non-linear term of the GPE depend on the azimuthal angle $\varphi $ and
thus $\hat{L}_\mathrm{z}$ and the operator $\hat{H}_{0}+NU_{0}|\psi
(\bi{r})|^{2}$ of the GPE commute, so that $m_\mathrm{c}$ is a good quantum
number. Since no further separation is possible, the equation must be solved in
the two variables $\rho $ and $z$. The ground state solution of the GPE
equation corresponds to $m_\mathrm{c}=0$.

Correspondingly the normal modes of the BdG equations will also have specific
angular momentum compositions if $\psi (\bi{r})$ is given by
(\ref{separation}). If $u_{i}(\bi{r})$ is an eigenfunction of
$\hat{L}_{\mathrm{z}}$ with eigenvalue $\hbar m$, then $v_{i}(\bi{r})$ will be
an eigenfunction with eigenvalue $\hbar(m-2m_\mathrm{c})$ \cite{Dodd97}.
Excitations on the ground state ($m_\mathrm{c}=0$) with $\pm |m|$ are
degenerate because $m$ enters quadratically into the BdG equations in
cylindrical coordinates.

The axially symmetric trap potential also has a reflection symmetry with
respect to the $x$-$y$ plane, and thus the solutions to the BdG equations can be
chosen to have a well-defined parity \cite{Hutchinson98}. In the isotropic
case, where $l$ and $m$ are good quantum numbers for the excitations, the
parity is simply given by $\Pi =(-1)^{l-m}$. Hutchinson and Zaremba showed
that, as the trap geometry is altered to an anisotropic one, the degenerate
modes split into branches with different excitation frequencies, which emerge
continously from the isotropic case (see figure 1 in reference
\cite{Hutchinson98}). This allowed them to associate a number $l$ with each
branch depending on where it originates in the isotropic limit, even though
total angular momentum is no longer a good quantum number.

\section{Classification of Excitations on the Ground State}\label{sec:families}

\subsection{Thomas--Fermi limit\label{Fliesser}}

In the Thomas--Fermi limit a complete separation of variables is possible.
Stringari \cite{Stringari96} has shown that by casting the GPE into
hydrodynamic form and then taking the Thomas--Fermi limit in the hydrodynamic
regime (i.e.\ energies much smaller than the chemical potential) a second order
wave equation can be derived to describe the elementary excitations. Fliesser
\emph{et al}\ \cite{Fliesser97} recognized some underlying symmetries of this
equation by identifying three operators which commute with each other, and
introduced three corresponding quantum numbers $(n,j,m)$ that classify the
solutions completely. An explicit separation of the wave equation was achieved
in cylindrical elliptical coordinates $\xi$,  $\eta$ and $\varphi$, and in
terms of these variables the quantum numbers represent

\vspace{1ex} 
  {\centering    
  \begin{tabular}{cl}
     $n$: & order of polynomial in $\xi $ and $\eta $ \\ 
     $j$: & index to label different eigenvalues for fixed $n$ and $|m|\,;$ \\ 
          &$j$ runs from $0$ to $N=1+\mathrm{int}\left[ \frac{n}{2}\right] $ \\ 
     $m$: & $z$-component of angular momentum.%
  \end{tabular}}
\vspace{1ex}

The Thomas--Fermi approximation is known to be very accurate as a limiting case
of the full solution if $Na/r_{0}\gg 1$, where $r_{0}$ is the harmonic
oscillator length. In particular, the shape of the ground state wave function
is very well approximated. Although the solutions of the full BdG equations do
not strictly conserve these quantum numbers, we find that they exhibit in
general the same patterns and symmetries, and we will show, in the appropriate
regime, how the family classification scheme we develop in this paper can be
related to $n,j,m$.

\subsection{Families in the isotropic case}

We first consider the isotropic case because then the patterns of the
different mode families can be easily described in terms of Legendre
polynomials. As the trap geometry is changed from spherical to cylindrical
symmetry these patterns are continously modified, being squeezed in the
direction of the stronger confinement, but the basic character remains
recognizable.

In cylindrical coordinates the solutions for the excitations in an isotropic
trap, $u(\bi{r})=u_\mathrm{r}(r)Y_{lm}(\vartheta,\varphi )$, can also be
separated as $u(\bi{r})=u(\rho ,z)\rme^{\rmi m\varphi }$. The relation  
\begin{equation}
   Y_{lm}(\vartheta ,\varphi )=\frac{1}{\sqrt{2\pi }}\sqrt{\frac{2l+1}{2}
   \frac{(l-m)!}{(l+m)!}}P_{lm}(\cos \vartheta )\rme^{\rmi m\varphi }
\end{equation}
between the spherical harmonics $Y_{lm}(\vartheta ,\varphi )$ and the Legendre
polynomials $P_{lm}(\cos \vartheta )$ shows that $u(\rho ,z)$ is essentially
the radial function $u_\mathrm{r}(r)$ modulated by the Legendre polynomial 
$P_{lm}(\cos \vartheta )$, where $\cos \vartheta =z(\rho ^{2}+z^{2})^{-1/2}$ in
cylindrical coordinates. In fact, it turns out that the general shape of the
families is determined by the symmetries of the Legendre polynomials. We now
show that the family classification suggested by Hutchinson and Zaremba can be
generalized, in the isotropic case, as follows. First we assign a 
\textit{principal family number} which is given by
\begin{equation}
  f=l-|m|+1,  
  \label{famnum}
\end{equation}
and then an additional number characterizing the radial function is needed to
complete the classification into families. We shall introduce the \textit{nodal
family number} $n_\mathrm{r}$ for this purpose, which in the isotropic case is 
simply the number of nodes in the radial function. The \textit{family}
is  given by the pair ($f,n_\mathrm{r}$), which together with the
magnetic quantum number $m$ uniquely specifies any mode. In section
\ref{sec:aniso} we will show how this family classification generalizes to the
anisotropic case.

\begin{figure}[tph]
  \begin{center}
  \includegraphics[width=\textwidth]{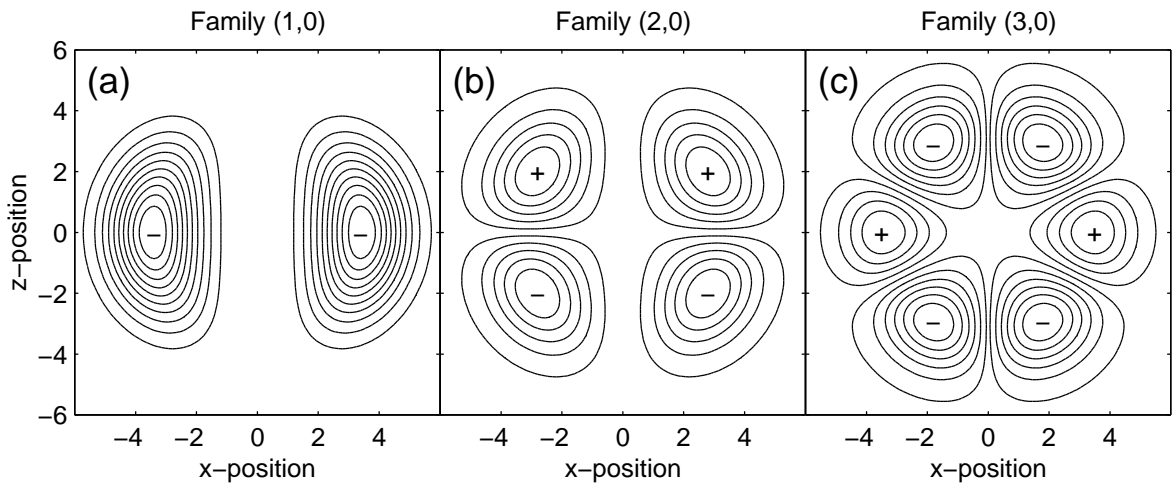}
  \end{center}
  \caption{General shape of mode families 1 to 3 with no radial node. Contour 
    plots in the $x$-$z$ plane of the quasiparticle amplitude $u(\rho,z)$ are 
    given for the degenerate modes $l=3$, (a) $m=3$, (b) $m=2$, (c) $m=1$ 
    modes.}
  \label{fig:families}
\end{figure}

We illustrate the spatial character of the family assignment by
considering first the excitation modes with no radial node ($n_\mathrm{r}=0$).
We begin with the case $m\neq0$, which we illustrate in figure
\ref{fig:families} with 
contour plots of full numerical solutions of $u(\rho,z)$ for the specific case
of the degenerate
modes of the lowest $l=3$ excitation with $m=3,2,1$. 
Their principal family numbers are $f=1,2,3$ respectively. The cylindrical
symmetry means that the $\rho$-$z$ dependence can be found on any plane through
the $z$-axis 
(we have chosen the $x$-$z$ plane), and the modes are symmetric
with respect to the $x$-coordinate. Since the radial function is the same for
each of these modes, the relative overall shape is determined by the Legendre
polynomials. The important property of the Legendre polynomials $P_{lm}(\cos
\vartheta )$ for our purposes is that they  have $n_{\vartheta}=l-|m|$ nodes
between $0 < \vartheta < \pi$. Thus, the number of \emph{angular} nodal
surfaces $n_{\vartheta}$ beween $0<\vartheta<\pi$, i.e. surfaces of zero
density that are characterized by a constant value of $\vartheta$ in the
isotropic case, determines the principal family number $f$ since $f$ is given
by equation (\ref{famnum}) as $f=l-|m|+1=n_{\vartheta}+1$. We note also that
for $m\neq0$ all Legendre polynomials are zero along the $z$-axis, but this is
not a nodal surface. Because the sign of the wave function changes as it
crosses a nodal surface, family 1 members have even parity, family 2 have odd
parity, and in general the parity of the mode is related to the principal
family number by $\Pi =(-1)^{f-1}$.

\begin{figure}[tph] 
  \begin{center}
    \includegraphics{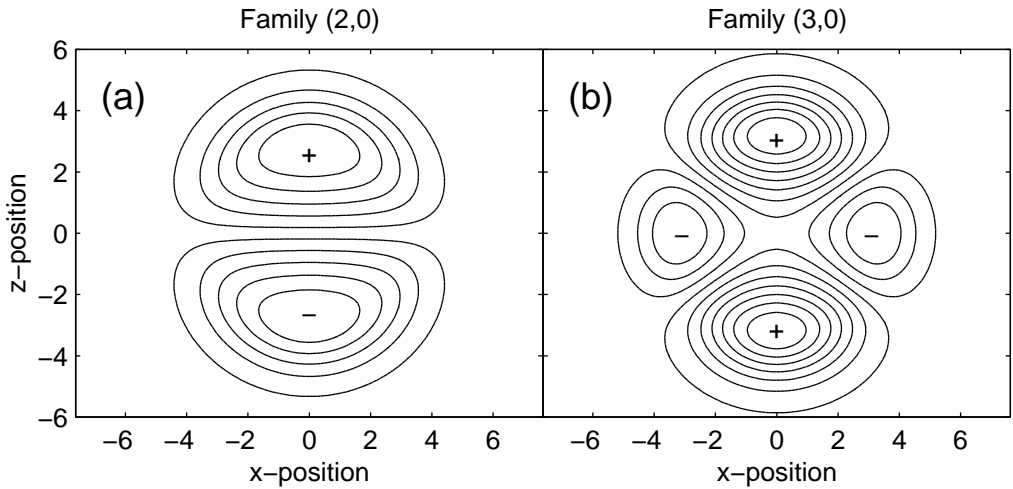} 
  \end{center} 
  \caption{Contour plots in the $x$-$z$ plane of the amplitude $u(\rho,z)$ of 
    the anomalous first members of family 2 and 3 ($m=0$).} 
  \label{fig:famm0} 
\end{figure}

The $m=0$ member of each family has a shape that derives from the $P_{l0}$
Legendre polynomial. We call it the \emph{anomalous} member of the family,
since its shape differs from other members of the family only in that it is
non-zero along the symmetry axis, which does not change the character of the
excitation significantly. We illustrate the shape of the anomalous modes of the
families $(2,0)$ and $(3,0)$ in figure \ref{fig:famm0}. 
We note also that the anomalous member of family $(1,0)$ is
the ground state, which is a solution of the BdG equations \cite{Morgan98}.

\begin{figure}[tph]
  \begin{center}
    \includegraphics[width=\textwidth]{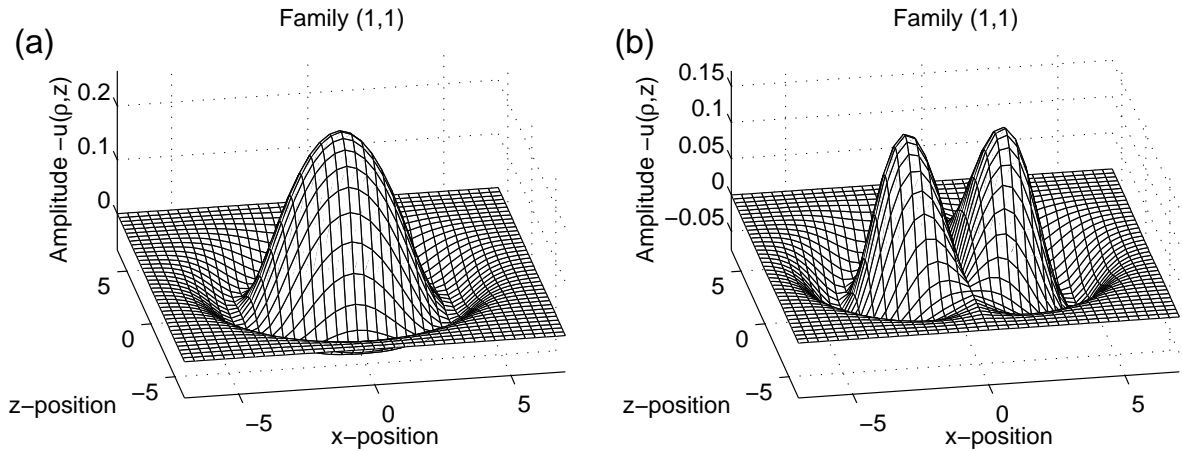}
  \end{center}
  \caption{Family 1 with one radial node ($n_\mathrm{r}=1$). (a) Anomalous
    first member ($l=0$, $m=0$), (b) general shape (here $l=1$, $m=1$).}
  \label{fig:fam2}
\end{figure}

The case where the radial function has a non-zero number of nodes (i.e.\ 
$n_{\mathrm{r}}\neq 0$) can now be easily visualized. The principal family
number $f$ determines the number of angular nodal surfaces $(f-1)$  between
$0<\vartheta <\pi$, while $n_\mathrm{r}$ determines the number of
\textit{radial} nodal surfaces, which intersect the angular nodal surfaces.
In the isotropic case they are spherical and centered on the origin. In figure
\ref{fig:fam2} we illustrate the first two modes having one node in the radial
function, 
which both belong to the family $(1,1)$. The mode in figure \ref{fig:fam2} (a)
is the anomalous first member of this family ($m=0$), while all other modes of
this family have the general shape shown in figure \ref{fig:fam2} (b), which
can be recognized as the same shape as in figure \ref{fig:families} (a), but
with one radial node. 

We stress that all members of the same family, apart from the anomalous one,
have the same general shape, i.e.\ the same number of peaks (in the contour
plot) in similar spatial distribution. The main qualitative difference between
modes of the same family is that the peaks move radially outwards and become
narrower in both radial and azimuthal direction with increasing eigenfrequency.
We also note that in the isotropic case the principal family number $f$ can be
readily obtained from the contour plots by counting the peaks in the half-plane
$x\ge0$ if there is no radial node ($n_\mathrm{r}=0$), or counting the peaks in
the region that is enclosed by the $z$-axis and the $n_\mathrm{r}=1$ radial
nodal surface if $n_\mathrm{r}>0$. For $m=0$ all peaks centred on the $z$-axis
must also be included.

To illustrate the mode classification by family and $m$ value, we list in
table \ref{tab:fam} the 18 lowest lying modes for the $C=332$ case in an
isotropic harmonic trap.
\begin{table}[tph]
\caption{Lowest quasiparticle modes of a condensate in an isotropic trap for 
$C=332$, listed by family.}
\begin{indented}
  \item[]
    \begin{minipage}[t]{5.5cm}
      \begin{tabular}[t]{ccc p{.3em} cc}
      \br
      \centre{3}{Mode}     & & \centre{2}{Family} \\
      $l$ & $m$ & $\omega$ & & $f$ & $n_\mathrm{r}$ \\
      \mr \ns
      0 & 0 & 0.000 & & 1 & 0\\
      \hline
      1 & 0 & 1.000 & & 2 & 0\\
        & 1 &       & & 1 & 0\\
      \hline
      2 & 0 & 1.545 & & 3 & 0\\
        & 1 &       & & 2 & 0\\
        & 2 &       & & 1 & 0\\
      \hline
      3 & 0 & 2.115 & & 4 & 0\\
        & 1 &       & & 3 & 0\\
        & 2 &       & & 2 & 0\\
        & 3 &       & & 1 & 0\\
      \br
      \end{tabular}
    \end{minipage}
    \begin{minipage}[t]{5.5cm}
      \begin{tabular}[t]{ccc p{2mm} cc}
      \br
      \centre{3}{Mode}     & & \centre{2}{Family} \\
      $l$ & $m$ & $\omega$ & & $f$ & $n_\mathrm{r}$ \\
      \mr \ns
      0 & 0 & 2.187 & & 1 & 1\\
      \hline
      4 & 0 & 2.748 & & 5 & 0\\
        & 1 &       & & 4 & 0\\
        & 2 &       & & 3 & 0\\
        & 3 &       & & 2 & 0\\
        & 4 &       & & 1 & 0\\
      \hline
      1 & 0 & 2.879 & & 2 & 1\\
        & 1 &       & & 1 & 1\\
      \br
      \end{tabular}
    \end{minipage}
  \end{indented}
\label{tab:fam}
\end{table}

\subsection{Anisotropic case}\label{sec:aniso}

The main value of the concept of families is in its extension to the
anisotropic cylindrically symmetric case. Hutchinson and Zaremba identified
the first four families by the dependence of the eigenvalue on trap
anisotropy. Here, we show that the mode topology determines the family.

\begin{figure}[th]
  \begin{center}
  \includegraphics[width=\textwidth]{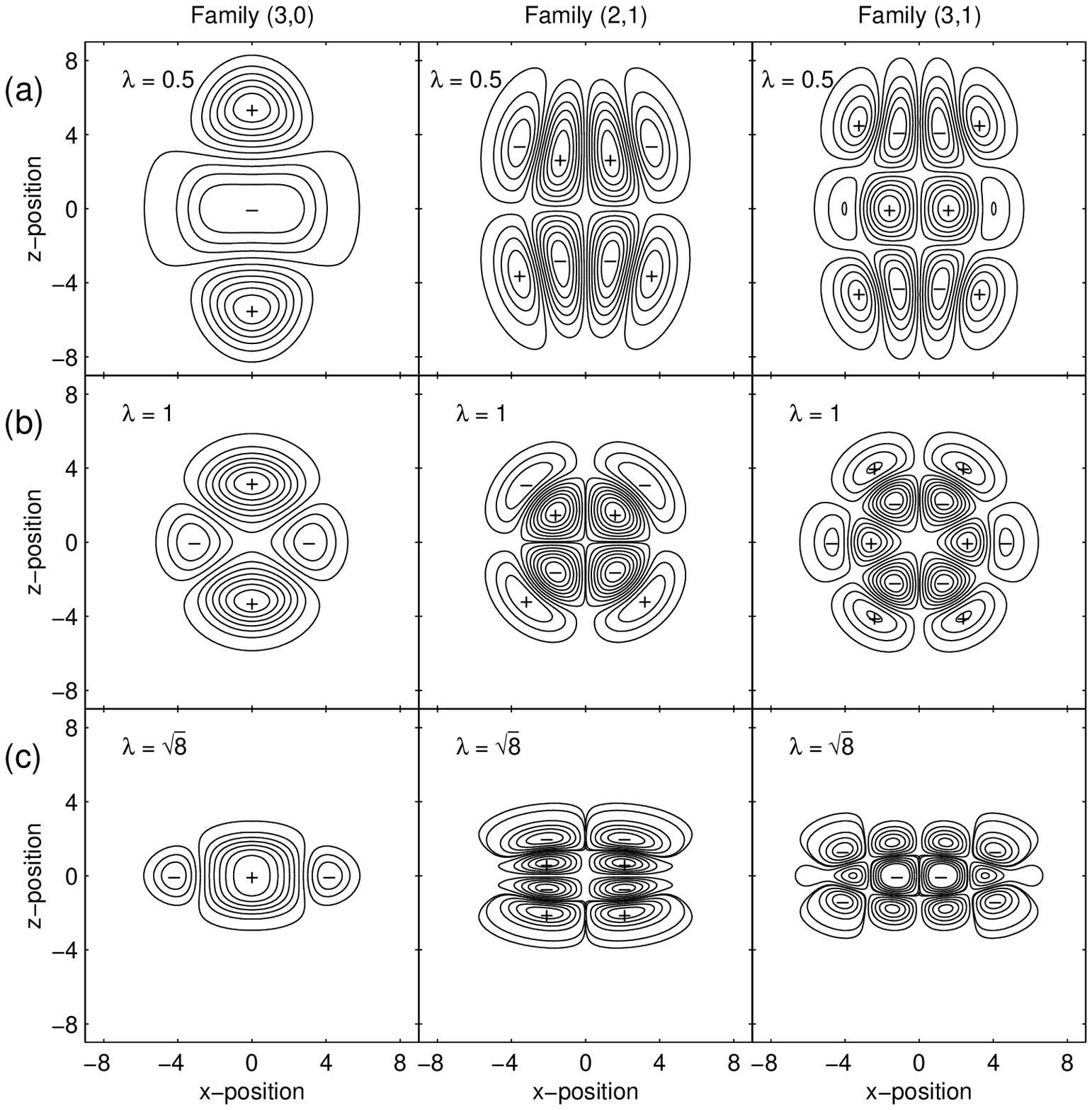}
  \end{center}
  \caption{Effects of trap anisotropy on quasiparticle shapes. Left hand column
    $f=3$, $n_\mathrm{r}=0$, $m=0$ mode; middle column $f=2$, $n_\mathrm{r}=1$, 
    $m=1$ mode; right hand column $f=3$, $n_\mathrm{r}=1$, $m=1$ mode. The 
    anisotropy parameter is (a) $\protect\lambda =0.5$, 
    (b) $\protect\lambda=1$, (c) $\protect\lambda =\protect\sqrt{8}$.}
  \label{fig:aniso}
\end{figure}

In figure \ref{fig:aniso} we have plotted the quasiparticle wave functions for
three families: the anomalous member of the ($3,0$) family, and the ($2,1$) and
($3,1$) families, for the case of prolate, spherical and oblate traps. 
These graphs illustrate the general features that are found in the anisotropic
case of cylindrical symmetry. We see that just as for the ground state of the
GPE the quasiparticles are squeezed in the direction of the stronger
confinement of the trap and expand in the other direction. For $\lambda <1$
(prolate) they are compressed in the $\rho$-direction and for $\lambda >1$
(oblate) they are squeezed in the $z$-direction. This distortion has no effect
on the zero line along the symmetry axis, but the nodal surfaces of the
isotropic case are distorted. The radial nodal surfaces are no longer
spherical, but are changed to a shape that approximately follows the
equipotentials of the trap. In all cases the total number of nodal crossings
along the  positive half of the axis of strong confinement  remains exactly
$n_{\mathrm{r}}$. The angular nodal surfaces can no longer be parameterized by
constant $\vartheta$, and they do not meet at the origin: in the prolate case
they generally intersect the symmetry axis at various (non-zero) values of $z$,
while in the oblate case they generally intersect the $z=0$ plane in circles of
different radii. We will call the former \textit{planar} nodal surfaces and the
latter \textit{cylindrical} nodal surfaces, which is a good description in
highly anisotropic traps as we will see in section \ref{sec:BdG}. Exceptions
are rare and occur for the case where the trap is close to spherical. Then, it
is possible in the prolate case that adjacent planar nodal surfaces join just
before meeting the $z$-axis, while in the oblate case it is possible that
adjacent cylindrical nodal surfaces join just before the $z=0$ plane. 

The nodal crossings of the isotropic case, which are at the centre of four
peaks of alternating sign, can also change character in the anisotropic case,
as illustrated in the right hand column of figure \ref{fig:aniso} (a). There we
see that pairs of peaks of the same sign can begin fusing; in other words the
crossing has become a saddle point, or `\textit{anti-crossing}'.  Despite these
distortions, the character of the isotropic excitations remains clearly evident
in the anisotropic case, and thus a family assignment can be made. The first
step is to determine the number of radial nodes (by inspection along the
positive part of the axis of strong confinement), which gives the radial family
number $n_\mathrm{r}$. Next, if there is no radial node  (i.e.
$n_{\mathrm{r}}=0$), we count the number $n_\mathrm{p}$ of  distinct peaks in
a  single quadrant (including peaks that are centered on the axes). If a radial
node does exist,  we identify the first radial nodal surface, and count the
number $n_\mathrm{p}$ of distinct peaks in a single quadrant inside this radial
nodal surface. In either case, the principal family number is then given by
$f=2n_\mathrm{p}-1$ for even modes and $f=2n_\mathrm{p}$ for odd modes. The
$n_{\mathrm{r}}=1$ nodal surface can be traced out by following a path from the
initial point on the axis of strong confinement, through successive crossings
or anti-crossings, until the axis of weak confinement is reached.  When 
anti-crossings are encountered, the path continues over the saddle (along the
line of minimum amplitude) to the nodal line leaving the anti-crossing opposite
to the entry. Of course this line (or radial surface) is no longer strictly
nodal, but nevertheless serves to determine a  region which is characteristic
of the family. Just as for the isotropic case,  a mode is uniquely identified
when the family assignment $(f,n_{\mathrm{r}})$ is given together with the
magnetic quantum number $m$, and the  parity is given by $\Pi =(-1)^{f-1}$.

\section{Ordering of Quasiparticle Eigenfrequencies}

In this section we show how the family classification can be related to the
energy ordering of the quasiparticles for given $m$, and we provide a simple
model to explain the relationship.

\subsection{Full solutions of Bogoliubov-de Gennes equations}\label{sec:BdG}

\begin{figure}[htp!]
  \begin{center} 
    \parbox{.49\textwidth}
      {\includegraphics[scale=.9]{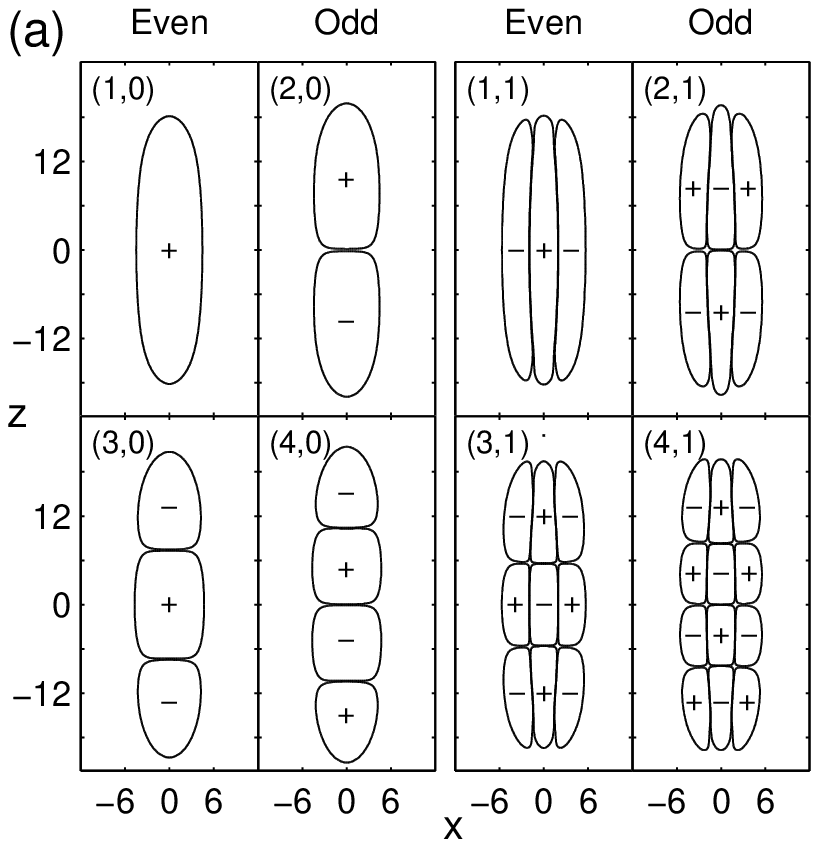}}
    \hspace{.002\textwidth}
    \parbox{.49\textwidth}
      {\includegraphics[scale=.9]{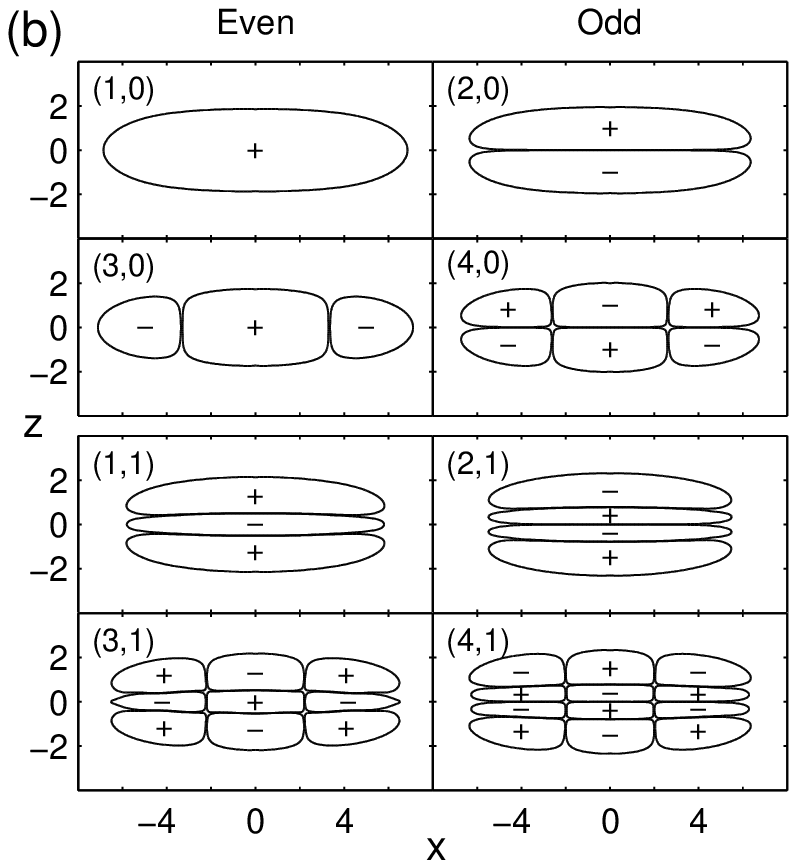}}
  \end{center}
  \caption{Highly anisotropic quasiparticle eigenstates in (a) prolate trap 
    with $1/\lambda=7$ and (b) oblate trap with $\lambda=7$. The cases shown 
    are the $f=1,2,3,4$ anomalous ($m=0$) family members with $n_\mathrm{r}=0$ 
    and 1, respectively. The contour plots show contour lines only at 
    $\pm 0.001$.} 
  \label{fig:fams}
\end{figure}

In figure \ref{fig:fams} we illustrate the shape of the first few family
members with $n_{\mathrm{r}}=0$ and $1,$ for the case of very prolate and
very oblate traps (figures \ref{fig:fams} (a) and (b) respectively). 
We have chosen to represent the $m=0$ modes, because the mode shapes are
slightly simpler than the $m\neq 0$ modes (which differ only in having a zero
along the $z$-axis). In highly anisotropic traps, the family shapes follow very
well defined patterns, as we can see, and it is convenient to discuss the
prolate and oblate cases separately.

\paragraph{Prolate case:}

\begin{figure}[htp!]
  \begin{center}
    \includegraphics[width=\textwidth]{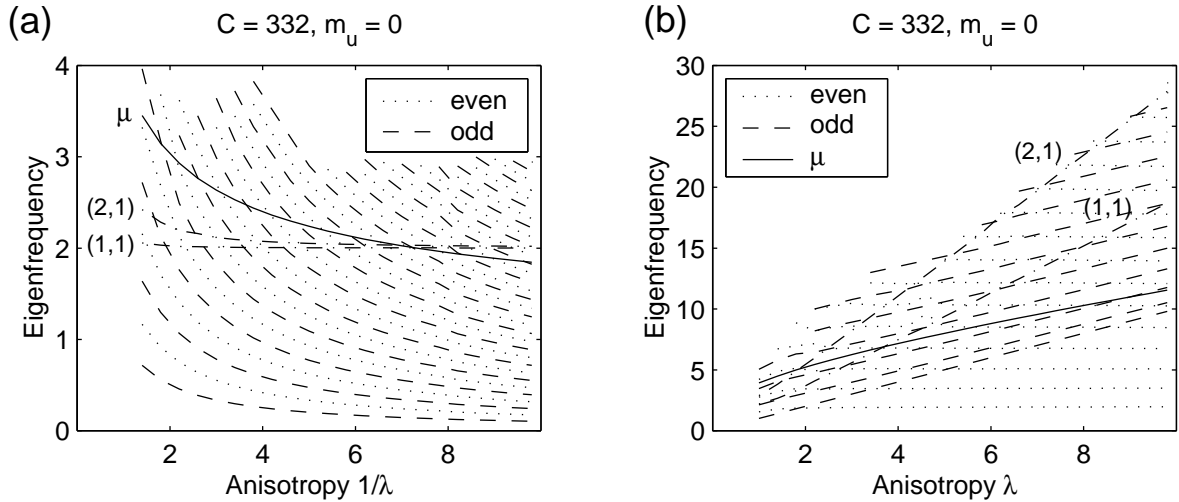}
  \end{center}
  \caption{Frequency spectrum versus anisotropy of the low-lying 
    $n_\mathrm{r}=0$ modes and the first two $n_\mathrm{r}=1$ modes ($m=0$) in 
    (a) prolate, (b) oblate geometry. The even and odd modes with 
    $n_\mathrm{r}=0$ are shown as dotted and dashed lines, respectively, and 
    the $n_r=1$ modes as dashed-dotted lines. The solid line shows the 
    condensate chemical potential $\mu$. The mode eigenfrequencies are measured 
    relative to the condensate eigenvalue $\mu$.}
  \label{fig:spectrum}
\end{figure}

For the prolate case an increase in the principal family number $f$ simply adds
an additional planar nodal surface perpendicular to the $z$-axis.
Changing $n_{\mathrm{r}}$ from 0 to 1 adds a radial nodal surface which
appears in this projection as two lines symmetric about, and almost parallel to
the $z$-axis. The energy ordering of these modes is plotted in figure
\ref{fig:spectrum} (a)  where we see that, for a given anisotropy, the energy
ordering initially follows the family assignment $(f,0)$ with alternating even
and odd modes. 
This sequence is interrupted by the $(1,1)$ mode, at an energy near 2, where it
has become more favourable to have a single radial nodal surface than many
planar ones across the narrow dimension. We note that the higher the anisotropy
(i.e.\ the larger $1/\lambda$) the more ($f,0$) modes fit in before the $(1,1)$
mode becomes favourable.

\paragraph{Oblate case:}

Figure \ref{fig:fams} (b) shows that in the oblate case it is natural to
separate the sequences of even and odd modes. Then, successive members of each
sequence (for fixed $n_{\mathrm{r}})$ are obtained by adding an additional
cylindrical nodal surface about the $z$-axis, which in this projection appears 
as a pair of nodal lines parallel and symmetrically displaced with respect to 
the $z$-axis. Changing  $n_{\mathrm{r}}$ from 0 to 1 adds again a radial nodal
surface, which appears here as a pair of nodal lines parallel to and 
symmetrically placed about the $z=0$ plane.

In figure \ref{fig:spectrum} (b), where the energies of the oblate modes are
plotted as a function of the anisotropy $\lambda$, the distinction between the
even and odd modes is clearly revealed. The odd modes are shifted as a group to
higher energies than the even modes (reflecting the energy cost of the nodal
surface through $z=0$), so that the energy ordering at a given anisotropy no
longer simply alternates between even and odd. We see too that the even modes
have essentially constant spacing except for very low $\lambda$, and the
spacing of the odd modes, while compressed for low excitation numbers, becomes
equal to the spacing of the even modes for higher excitation numbers. Once
again, as for the prolate case, the higher the anisotropy the more $(f,0)$
modes fit in before the first  $n_{\mathrm{r}}=1$ mode.

\subsection{Comparison with harmonic oscillator solutions}

\label{sec:HO}

For higher excitations the quasiparticle amplitude $v_{i}$ is negligible
compared to $u_{i}$ and the BdG equations (\ref{BdG}) reduce to the
eigenvalue problem 
\begin{equation}
  [\hat{H}_{0}+2NU_{0}|\psi_{\mathrm{g}}(\mathbf{r})|^{2}-\mu ]u_{i}(\mathbf{r})
   = \hbar \omega _{i}u_{i}(\mathbf{r}),
  \label{reducedBdG}
\end{equation}
which has the form of a single particle equation \cite{You97,Dalfovo97}. The
presence of the condensate gives an effective potential that is broadly
harmonic, but with a repulsive dimple in the middle, of height approximately
$\mu $. At high excitations, the energy levels for equation (\ref{reducedBdG})
will be essentially those of the harmonic oscillator, while for mode energies
comparable or less than $\mu ,$ the presence of the dimple will shift the
energy levels upwards, effectively compressing them.

If the condensate density is ignored in equation (\ref{reducedBdG}) the
equation is simply that of the harmonic oscillator and is separable in
cylindrical coordinates. The energy levels are given in units of $\hbar 
\omega_{\mathrm{r}}$ by 
\begin{eqnarray}
  E=|m|+1+2n_{\rho }+2\lambda (n_{\mathrm{z}}+\frac{1}{4})\qquad  &&
    \mbox{(even modes),}  \label{HOeven} \\
  E=|m|+1+2n_{\rho }+2\lambda (n_{\mathrm{z}}+\frac{3}{4}) &&
     \mbox{(odd modes),}  \label{HOodd}
\end{eqnarray}
where $n_{\rho }$ and $n_{\mathrm{z}}$ are the number of nodes in the $\rho$- 
and $z$-wave function respectively ($n_{\rho }=0,1,\ldots $, 
$n_{\mathrm{z}}=0,1,\ldots $). Equations (\ref{HOeven}) and (\ref{HOodd})
provide considerable insight into the behaviour of the solutions we have
obtained from the full BdG equations.

\begin{figure}[htp!]
  \begin{center}
    \includegraphics{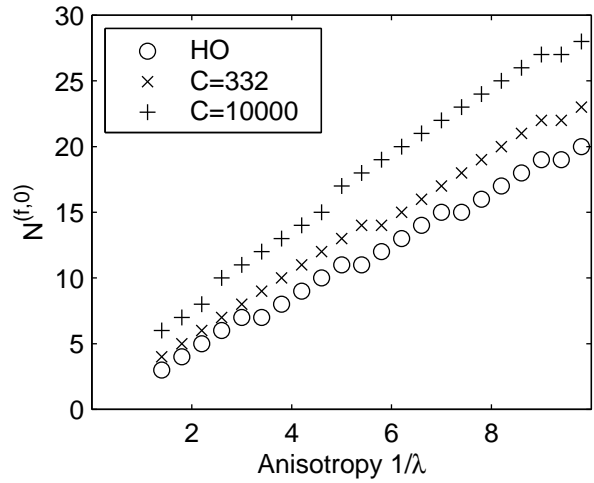}
  \end{center}
  \caption{Comparison of energy ordering predictions of full BdG and harmonic
    oscillator solutions in a \emph{prolate} trap. Plotted points show 
    $N^{(f,0)}$, the number of $n_\mathrm{r}=0$ modes with energy lower than 
    the $(1,1)$ mode. All modes are $m=0$.}
  \label{fig:HOcompprol}
\end{figure}

In the prolate case, increasing $n_{\mathrm{z}}$ corresponds to increasing $f$
in the BdG solutions,  while increasing $n_{\rho }$ corresponds to increasing
$n_{\mathrm{r}}$ in the BdG solutions. For $\lambda <1$ it becomes
energetically favourable to increase $n_{\mathrm{z}}$ rather than $n_{\rho }$,
and we can fit roughly $1/\lambda$ nodes in the  $z$-direction before it is
energetically favourable to put a node in the  $\rho$-direction. In figure
\ref{fig:HOcompprol} we show from calculations of the BdG equations the number 
$N^{(f,0)}$ of $(f,0)$ modes of lower energy than the first $n_{\mathrm{r}}=1$
mode and compare this qualitatively to the  predictions from equations
(\ref{HOeven}) and (\ref{HOodd}).
We see that while there is relatively good agreement at $C=332$, the full BdG
calculation predicts more $(f,0)$ modes to fit in, and the discrepancy
increases at higher values of $C$. This is due to the fact that in the full
equations the  presence of the condensate causes the spacing of the lower
levels (up to an energy of order $\mu$ above the condensate) to be  compressed,
and  $\mu$ increases with $C$.  

\begin{figure}[htp!]
  \begin{center}
  \includegraphics[width=\textwidth]{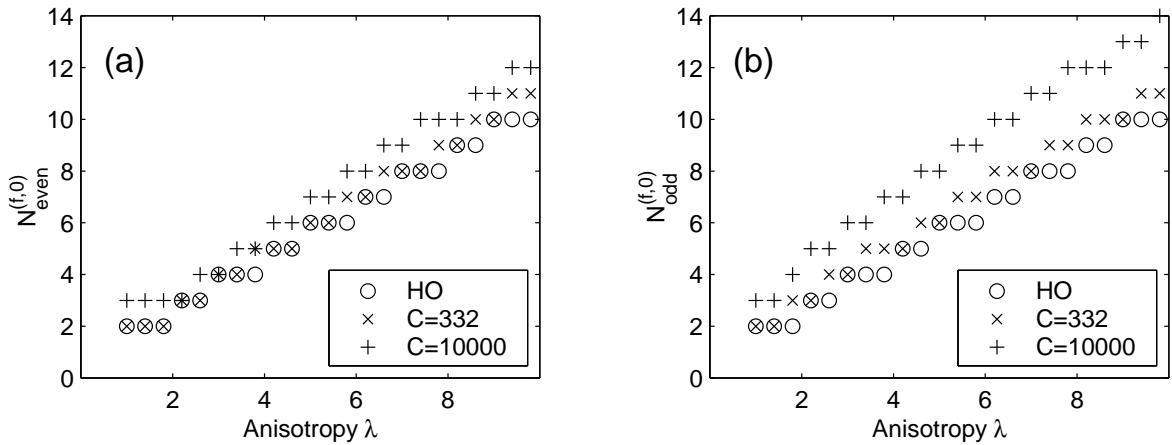}
  \end{center}
  \caption{Comparison of energy ordering predictions of full BdG and harmonic
    oscillator solutions in the \emph{oblate} case for (a) the number 
    $N^{(f,0)}_{\mathrm{even}}$ of even $n_\mathrm{r}=0$ modes with energy 
    lower than $(1,1)$ mode, (b) the number $N^{(f,0)}_{\mathrm{odd}}$ of odd 
    modes with energy lower than the (2,1) mode. All modes are $m=0$.}
  \label{fig:HOcompobl}
\end{figure}

In the oblate case, the agreement between the harmonic oscillator solutions and
the BdG solutions is much better. As explained in the previous section, it is
useful to separate the even modes (where  $n_{\mathrm{z}}$ corresponds to
$2n_{\mathrm{r}}$, and $n_{\rho}$ to $(f-1)/2)$ and the odd modes (where
$n_{\mathrm{z}}$ corresponds to $2n_{\mathrm{r}}+1$, and $n_{\rho }$ to
$(f-2)/2)$. Since  $\lambda >1$, it is now clear from equations (\ref{HOeven})
and (\ref{HOodd}) that  increasing  $n_{\mathrm{z}}$, i.e.\ increasing
$n_{\mathrm{r}}$ in the BdG solutions, is more energetically costly than
increasing $n_\rho$, which corresponds to increasing $f$. Figure
\ref{fig:HOcompobl} provides a quantitative comparison of the predictions of
the harmonic oscillator solutions and the BdG solutions for the number of
$(f,0)$ modes of lower energy than the first $n_{\mathrm{r}}=1$ mode.  
We have separated the even and odd modes so that figure \ref{fig:HOcompobl} (a)
shows the number $N^{(f,0)}_{\mathrm{even}}$ of even modes with eigenfrequency
lower than that of the (1,1) mode, while figure \ref{fig:HOcompobl} (b) shows
the number $N^{(f,0)}_{\mathrm{odd}}$ of odd modes with eigenfrequency lower
than that of the (2,1) mode. The agreement is excellent for the $C=332$ case
and still very good for the $C=10000$ case and  is much better than for the
prolate case of corresponding asymmetry, because  the chemical potential $\mu $
lies lower relative to the eigenfrequencies in question, as can be seen in
figure \ref{fig:spectrum}.

Finally, we remark that the magnetic quantum number $m$ simply gives an energy
offset in equations (\ref{HOeven}) and (\ref{HOodd}), i.e.\ modes with a given
number of nodes have higher energies for higher $m$. It is clear 
therefore, that the results discussed above for $m=0$ will also apply for 
$m\ne0$. The energy offset by $m$ also explains that for a given family 
$(f,n_{\mathrm{r}})$ (i.e. a given number of nodes) the energy increases with 
increasing $m$ (see table \ref{tab:fam}).

\section{Discussion}

We have shown that the concept of mode families introduced by Hutchinson
and Zaremba \cite{Hutchinson98} can be systematically defined and extended to
include all excitation modes of cylindrically symmetric anisotropic traps. The
family assignment $(f$, $n_\mathrm{r})$ determines the topology of any mode
with  $m\neq 0,$ and the $m=0$ mode differs only in being non-zero on the
symmetry axis. In the regime where the treatment of Fliesser \emph{et al}\
\cite{Fliesser97} is valid, we can relate their quantum numbers $n$, $j$ to our
family assignment numbers $f$ and $n_\mathrm{r}$ as follows:
\begin{equation}
  f=n-2j+1,\qquad n_\mathrm{r}=j.
\end{equation}

The similarity of shape of modes of the same family explains the similarity of 
frequency dependence on anisotropy for modes in a given family found by
Hutchinson and Zaremba; changes in trap geometry affect every mode in the
family in much the same way. We have also shown how the energy ordering of the
quasiparticle excitations is related to their family shape and provided a
simple model that explains the behaviour in terms of harmonic oscillator 
eigenstates.

\ack 
This work was supported by the Marsden Fund of New Zealand under Contract No.
PVT902.

\end{document}